# Transportable clock laser system with an instability of $1.6 \times 10^{-16}$

Sofia Herbers,[1] Sebastian Häfner,[1,2] Sören Dörscher,[1] Tim Lücke,[1] Uwe Sterr,[1] and Christian Lisdat[1,*]

[1]*Physikalisch-Technische Bundesanstalt, Bundesallee 100, 38116 Braunschweig, Germany*
[2]*Currently with Nanosystems and Technologies GmbH, Gleimershäuser Str. 10, 98617 Meiningen, Germany*
*Corresponding author: christian.lisdat@ptb.de

We present a transportable ultra-stable clock laser system based on a Fabry–Perot cavity with crystalline $Al_{0.92}Ga_{0.08}As/GaAs$ mirror coatings, fused silica (FS) mirror substrates, and a 20 cm-long ultra-low expansion (ULE) glass spacer with a predicted thermal noise floor of mod $\sigma_y = 7 \times 10^{-17}$ in modified Allan deviation at one second averaging time. The cavity has a cylindrical shape and is mounted at 10 points. Its measured sensitivity of the fractional frequency to acceleration for the three Cartesian directions are $2(1) \times 10^{-12}$ /(ms$^{-2}$), $3(3) \times 10^{-12}$ /(ms$^{-2}$), and $3(1) \times 10^{-12}$ /(ms$^{-2}$), which belong to the lowest acceleration sensitivities published for transportable systems. The laser system's instability reaches down to mod $\sigma_y = 1.6 \times 10^{-16}$  © 2022 Optica Publishing Group

https://doi.org/10.1364/OL.470984

Among other applications, ultra-stable lasers as local oscillators largely determine the performance of optical clocks. These clocks enable very accurate time and frequency measurements, are considered for the redefinition of the second [1], and offer many applications such as research in fundamental physics on Earth as well as in space [2,3] and relativistic geodesy [4]. All these applications require a very low fractional frequency uncertainty and instability of the clock. Today's best clocks reach systematic uncertainties of $9.4 \times 10^{-19}$ [5] and instabilities of $5 \times 10^{-17}/\sqrt{\tau}$ [6,7]. Thus, a statistical uncertainty of the order of $10^{-18}$ is reached within minutes of measurement time. So far, only laboratory-based clocks reach such levels of performance, whereas transportable clocks, which are required by space applications and relativistic geodesy, reach systematic uncertainties of mid $10^{-18}$ but instabilities of $1 \times 10^{-15}/\sqrt{\tau}$ at best [8–11], which results in measurement times of more than a week to reach a statistical uncertainty of $10^{-18}$. This is a show-stopper for fast and time-resolved measurements.

The inferior instability of transportable clocks results from the transportable lasers' lower coherence that, compared with stationary ones, increases the resolvable linewidth and thus the instability from quantum projection noise [12] and reduces the clock duty cycle which increases the instability caused by the Dick effect [13]. The best stationary system reaches a fractional frequency instability in modified Allan deviation of down to mod $\sigma_y = 4 \times 10^{-17}$ [14], while the best transportable system only reaches an instability of mod $\sigma_y = 3 \times 10^{-16}$ [15].

The vast majority of clock laser systems stabilize the laser's frequency to a resonance of an ultra-stable Fabry–Perot cavity using the Pound–Drever–Hall (PDH) technique. This can ultimately reduce the laser's fractional frequency instability to the fractional length stability of the ultra-stable cavity. The latter is fundamentally limited by optical path length fluctuations caused by thermal noise in the cavity components. While stationary systems profit from long spacers [16] or operating the optical cavity at cryogenic temperatures [14] to reduce the thermal noise floor, transportability imposes limitations on the length and operation temperature. Thus, transportable systems are operated around room temperature to have the system ready for operation in reasonable time after transport. New mirror materials with advantageous material properties such as crystalline $Al_{0.92}Ga_{0.08}As/GaAs$ mirror coatings [17] or meta-mirrors [18] are promising candidates to reduce the thermal noise floor.

Another limitation of the length stability of a cavity are vibrations affecting its geometric length. Stationary systems benefit from soft and acceleration-insensitive mountings, which allow for acceleration sensitivities of the fractional frequency down to few $10^{-12}$ /(ms$^{-2}$) [14,19]. The mounting of a transportable system must be acceleration insensitive, but at the same time sufficiently rugged for transportation. In the past years, transportable mounting designs have been developed (Table 1), which allow for acceleration sensitivity comparable with those of stationary systems. To outperform today's best transportable lasers, an acceleration sensitivity of no more than a few $10^{-11}$ /(ms$^{-2}$) is needed, assuming operation on commercial vibration isolation platforms, which reduce accelerations to the order of $10^{-5}$ ms$^{-2}$.

Furthermore, photothermal noise, pressure, and temperature fluctuations affect the cavity's optical path length. Residual amplitude modulation (RAM), which is an unavoidable by-product of the phase modulation required by the PDH method, and optical path length fluctuations from the clock laser to the ultra-stable cavity or to the atoms can degrade the stability transfer from the ultra-stable cavity to the atoms. Therefore, all these effects need to be controlled or suppressed to not increase the frequency instability of the clock laser's light at the atoms significantly above the thermal noise floor.

**Table 1. Transportable Cavity Designs with Acceleration Sensitivities of the Order of $10^{-11}$ /ms$^{-2}$ and Below[a]**

| Design | Acceleration Sensitivity | Reference |
| --- | --- | --- |
| Spherical 5 cm | $0.2 \times 10^{-11}$ /(ms$^{-2}$) | [20] |
| Cubic 5 cm | $0.1 \times 10^{-11}$ /(ms$^{-2}$) | [21] |
| Cubic 10 cm | $2 \times 10^{-11}$ /(ms$^{-2}$) | [22] |
| V. cylindrical 10 cm | $4 \times 10^{-11}$ /(ms$^{-2}$) | [23] |
| H. cylindrical 10 cm | $4 \times 10^{-11}$ /(ms$^{-2}$) | [24] |
| H. cylindrical 20 cm | $0.3(3) \times 10^{-11}$ /(ms$^{-2}$) | This work |

[a]Abbreviations are: vertical (V.); horizontal (H.).

Here, we present a laser system for a transportable $^{87}$Sr lattice clock [8] designed to surpass and replace the lowest instability transportable laser system with an instability down to mod $\sigma_y = 3 \times 10^{-16}$ reported so far [15]. The new laser system is based on crystalline Al$_{0.92}$Ga$_{0.08}$As/GaAs mirror coatings [17], fused silica (FS) mirror substrates, ultra-low expansion (ULE) glass compensation rings [25], and an ULE glass spacer with a length of 20 cm. These coatings are highly absorbing at visible wavelengths. Thus, the laser system is operated at 1397 nm, which is twice the interrogation wavelength of a $^{87}$Sr lattice clock. This setup allows for a thermal noise floor of approximately mod $\sigma_y = 7 \times 10^{-17}$ at one second, see Supplement 1.

Crystalline AlGaAs coatings exhibit birefringence [17,26,27]. Because the crystalline axis of the coatings was not marked correctly, the mirrors are not aligned to maximum mode splitting. Thus, we measured a relatively small mode splitting of 16 kHz, which is sufficient to separate the polarization eigenmodes.

A cavity finesse of approximately $3 \times 10^5$ was measured for both the TEM$_{00}$ mode and the TEM$_{10}$ mode within a test setup. However, after placing the cavity within its final setup the finesse reduced to $1.36(9) \times 10^5$ for the TEM$_{00}$ mode, while for the TEM$_{10}$ mode the finesse remained at $2.94(8) \times 10^5$. Most likely, a dust particle that fell onto the mirror reduced the finesse as the finesse of the TEM$_{01}$ mode has also reduced. Hence, the cavity is operated on the TEM$_{10}$ mode as its higher finesse reduces the impact of technical imperfections such as RAM on the fractional frequency instability.

To operate the optical clock and perform clock comparisons, the light of an external cavity diode laser is sent via optical fibers to the ultra-stable reference cavity, the atoms, and a second clock or a frequency comb. Optical path length fluctuations would limit the transfer at a fractional instability of $10^{-16}$. Therefore, a lightweight and compact laser distribution breadboard has been designed to include Doppler cancellation [28] for all three paths including the frequency doubling (Fig. S4a, Supplement 1). From beat note measurements between the different optical outputs of the distribution breadboard we derive the instability caused by the frequency transfer, which is below the thermal noise limit for Fourier frequencies below 2 Hz (Fig. 1).

To characterize the laser system, light was sent via the distribution breadboard to the ultra-stable cavity and to a frequency comb stabilized to a second ultra-stable laser system [14]. The laser is locked with the PDH method to the ultra-stable cavity, with a shot noise limit near $S_y = 10^{-35}$ Hz$^{-1}$ and an incident power of the order of 10 µW. Supplement 1 includes a detailed description of the setup.

We have investigated the instability caused by RAM for three electro-optic modulators (EOMs): EOM1 has a normal-incidence surface and an output surface tilted by 4° (PM7-SWWIR from QUBIG); EOM2 is a homemade Brewster-cut EOM with a lithium niobate (LiNbO$_3$) crystal, as used in our previous setups [15,16]; and EOM3 is a waveguide fiber-coupled EOM manufactured by EOSPACE. The free space EOMs and other optics needed for the PDH stabilization are placed on a breadboard, which is attached to the vacuum chamber of the cavity, (Fig. S5b, Supplement 1). For the Doppler compensation, the backreflection mirror is placed on the breadboard in front of the EOM. The path through the fiber-coupled EOM3 must be included in the Doppler compensation. However, light propagating bidirectionally through the EOM gives rise to étalon effects and degrades the laser stability by variable RAM. Hence, we used a two stage Doppler cancellation scheme that avoids light propagating back through the EOM [16].

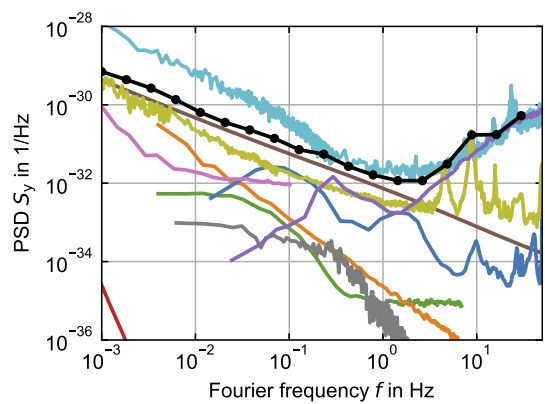

**Fig. 1.** Power spectral density (PSD) of fractional frequency fluctuation of the laser system (cyan) along with the expected instability (black), which is composed of the instability caused by thermal noise (brown), seismic noise (blue), thermally induced mechanical stress (pink), photothermal noise (orange), RAM (green), stability transfer (purple), pressure fluctuations (gray), temperature fluctuations (red), and the reference laser system (olive) [14]. See Dataset 1, Ref. [30] for underlying values.

We evaluated the fractional frequency instability caused by RAM as described in Supplement 1. The instability mod $\sigma_y$ caused by RAM using EOM1 is of the order of $10^{-16}$ around 1 s averaging time. Even with an additional active RAM stabilization [29], we could not reduce the instability significantly [27]. For EOM2, the resulting instability is in the mid $10^{-17}$ range without any RAM stabilization, which is on the same level as the expected thermal noise floor. The fractional frequency instability mod $\sigma_y$ related to EOM3 including an active RAM stabilization is below $2 \times 10^{-17}$ and thus, well below the thermal noise limit (Fig. 1). We use EOM3 to further evaluate the system to ensure that no limitation is caused by RAM.

The photothermal effect denotes the optical length change of a cavity caused by intracavity power fluctuations and thus, fluctuations of the power absorbed by the two mirrors [31]. It separates into three relevant contributions: the substrate's photothermoelastic contribution; the coating's photothermoelastic and photothermorefractive contribution. The two coating contributions usually have opposite sign and thus, partly cancel. We measured the cavity's length response via the beat note with the stabilized frequency comb by applying a sinusoidal modulation of the order of µW to the light power entering the cavity via an acousto-optic modulator (AOM). We derived a modulation of 0.2 µW in absorbed power per mirror by comparing the measurement with the model [31] using the equations and values given in Supplement 1. The measurement shows a faster roll-off at higher frequencies compared with the model (Fig. 2). The slight difference between the model and measurements might be caused by the poor knowledge of material properties put into the model. Chalermsongsak *et al.* [33] obtained similar results for the photothermal transfer function of AlGaAs mirror coatings.

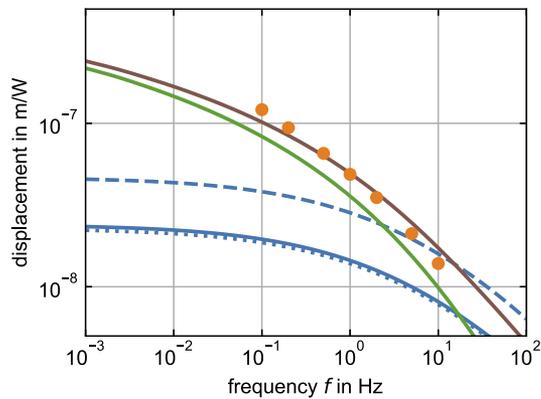

**Fig. 2.** Expected magnitude of the photothermal transfer function per watt (brown), which is composed of the photothermoelastic substrate contribution (green) and the coating contribution (solid blue), which is in turn composed of the photothermorefractive coating contribution (dotted blue) and the photothermoelastic coating contribution (dashed blue). Measured magnitude of the photothermal transfer function (orange) fitted to the expected function by adjusting the absorbed power per mirror. See Dataset 2, Ref. [32] for underlying values.

Without any further stabilization of the intracavity power, the instability resulting from the photothermal effect is of the order of mod $\sigma_y = 10^{-16}$. Thus, the intracavity power is stabilized using the output of a photodiode (PD) in transmission of the cavity and a proportional-integral controller that controls the RF drive power of an AOM. Power fluctuations in a beam coupled out with a beam sampler from of the incident light indicate that the expected instability caused by residual power fluctuations is below the thermal noise floor (Fig. 1).

Our mounting follows concepts as in Refs. [15,34], which provide independent, vibration insensitive mounting in the three translational degrees of freedom and avoid overdetermination in all six degrees of freedom. We implemented a simplified mount, which allows only for independent mounting in the three spatial directions. The cavity is fixed at 10 points with spring steel wires and a clamps made from fluorine rubber rings, washers, and nuts (Fig. 3). Compared with previous designs [16,24], our support allows for a soft mounting with low preload reducing the influence of tolerances on the acceleration sensitivity.

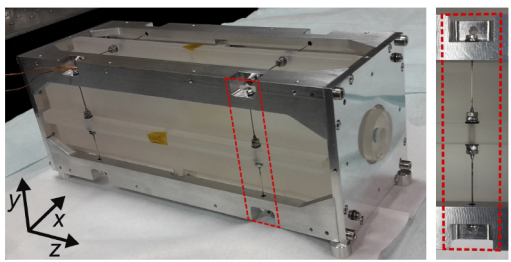

**Fig. 3.** Ultra-stable cavity within its mounting frame. Detail: spring steel wire welded to a threaded pin; fluorine rubber o-rings; washers, and nuts; see also Ref. [27].

In order to measure the cavity's acceleration sensitivity, we apply sinusoidal accelerations along different axes with an active vibration isolation platform. The acceleration is measured by a seismometer and the acceleration sensitivity is then derived from the modulations of the beat note with the frequency comb. We find acceleration sensitivities of $2(1) \times 10^{-12}$ /(ms$^{-2}$), $3(3) \times 10^{-12}$ /(ms$^{-2}$), and $3(1) \times 10^{-12}$ /(ms$^{-2}$) for the $x$-, $y$-, and $z$-axis, respectively. The acceleration sensitivity of this cavity belongs to the lowest sensitivities published so far for transportable cavities, even though the cavity is a up to four times longer than systems with a similar acceleration sensitivity (Table 1). Finally, we measured the residual acceleration on the vibration isolation platform to infer the instability caused mainly by seismic vibrations (Fig. 1).

The cavity including its mounting is enclosed by two passive heat shields, one active heat shield, and a vacuum chamber made of aluminum, (Fig. S6, Supplement 1). The temperature of the active heat shield is stabilized to the coefficient of thermal expansion (CTE) zero crossing temperature of approximately 295 K within 1 K. To evaluate the instability caused by temperature fluctuations (Fig. 1), we model the heat transfer between the heat shields as cascaded low passes [27,35].

After a temperature change of the active heat shield, an immediate change of the cavity length was observed via the beat note at the frequency comb (Fig. S7, Supplement 1), which could result from temperature-dependent mechanical stress in our setup, as the degrees of freedom are overdetermined in our mounting. Without a better model, we assume simple proportionality between fractional frequency change and the active heat shield's temperature and get a conversion factor of $3(2) \times 10^{-12}$ K$^{-1}$ (Fig. S7, Supplement 1). No such behavior is observed when changing the vacuum chamber's temperature. From the remaining temperature fluctuations of the outer heat shield and with the conversion factor we estimate the instability caused by mechanical stress (Fig. 1).

A change in pressure influences the refractive index of the residual gas between the mirrors. Using the refractive index of air and the measured pressure fluctuations this leads to an instability as shown in Fig. 1 [27,36]. The absolute pressure inside the vacuum chamber is of the order of $10^{-9}$ mbar.

The cyan line in Fig. 1 shows the laser system's measured instability and the black dots the total instability expected from all technical noise contributions. The measured instability matches the expected instability for Fourier frequencies above 0.2 Hz, while it is higher at lower frequencies. One explanation might be newly discovered noise sources in AlGaAs mirror coatings [37].

With an instability as low as mod $\sigma_y = 1.6 \times 10^{-16}$ (Fig. S3, Supplement 1), the instability of the laser system presented here already surpasses the instability of $2 \times 10^{-16}$ that would be possible using Ta$_2$O$_5$/SiO$_2$ mirrors at 1397 nm [27]. In direct comparison to its predecessor system [15], the instability of the laser system presented here is lower by factors of seven and 1.5 for averaging times below 0.1 s and above 1 s, respectively. This results in a two times larger coherence time. The clock instability caused by the Dick effect is expected to reduce by a factor of four for typical interrogation times below 1 s [27].

In conclusion, we have presented the transportable laser system with the lowest reported fractional frequency instability so far. The cavity's acceleration sensitivity is comparable with the lowest values of up to four times smaller cavities. To further reduce the laser system's instability, an active feed forward correction of residual accelerations to the laser's frequency [38] or improving the low frequency performance of the active vibration isolation platform [39] could be implemented to push down the instability to the thermal noise limit especially around 1 s of averaging time. The unexpected high instability for Fourier frequencies below 0.2 Hz needs further investigations especially in view of recently discovered new noise sources in AlGaAs mirror coatings [37].

**Funding.** European Metrology Programme for Innovation and Research (20FUN08 NEXTLASERS); Deutsche Forschungsgemeinschaft (EXC-2123 QuantumFrontiers (Project-ID 390837967), SFB 1464 TerraQ (Project-ID 263 434617780)).

**Acknowledgments.** We acknowledge support by the Project 20FUN08 NEXTLASERS, which has received funding from the EMPIR programme cofinanced by the Participating States and from the European Union's Horizon 2020 Research and Innovation Programme, and by the Deutsche Forschungsgemeinschaft (DFG, German Research Foundation) under Germany's Excellence Strategy–EXC-2123 QuantumFrontiers, Project-ID 390837967, and SFB 1464 TerraQ, Project-ID 434617780, within project A04.

**Disclosures.** US: Physikalisch-Technische Bundesanstalt (P).

**Data availability.** Data underlying the results presented in this paper are available in Dataset 1, Ref. [30], Dataset 2, Ref. [32], Dataset 3, Ref. [40], Dataset 4, Ref. [41], and Dataset 5, Ref. [42], respectively.

**Supplemental document.** See Supplement 1 for supporting content.

## REFERENCES

1. F. Riehle, P. Gill, F. Arias, and L. Robertsson, Metrologia **55**, 188 (2018).
2. P. Delva, A. Hees, and P. Wolf, Space Sci. Rev. **212**, 1385 (2017).
3. M. Safronova, D. Budker, D. DeMille, D. F. Jackson Kimball, A. Derevianko, and C. W. Clark, Rev. Mod. Phys. **90**, 025008 (2018).
4. T. Mehlstäubler, G. Grosche, C. Lisdat, P. Schmidt, and H. Denker, Rep. Prog. Phys. **81**, 064401 (2018).
5. S. M. Brewer, J.-S. Chen, A. M. Hankin, E. R. Clements, C. W. Chou, D. J. Wineland, D. B. Hume, and D. R. Leibrandt, Phys. Rev. Lett. **123**, 033201 (2019).
6. E. Oelker, R. B. Hutson, C. J. Kennedy, L. Sonderhouse, T. Bothwell, A. Goban, D. Kedar, C. Sanner, J. M. Robinson, G. E. Marti, D. G. Matei, T. Legero, M. Giunta, R. Holzwarth, F. Riehle, U. Sterr, and J. Ye, Nat. Photonics **13**, 714 (2019).
7. R. Schwarz, S. Dörscher, A. Al-Masoudi, E. Benkler, T. Legero, U. Sterr, S. Weyers, J. Rahm, B. Lipphardt, and C. Lisdat, Phys. Rev. Res. **2**, 033242 (2020).
8. S. B. Koller, J. Grotti, S. Vogt, A. Al-Masoudi, S. Dörscher, S. Häfner, U. Sterr, and C. Lisdat, Phys. Rev. Lett. **118**, 073601 (2017).
9. Y. Huang, H. Zhang, B. Zhang, Y. Hao, H. Guan, M. Zeng, Q. Chen, Y. Lin, Y. Wang, S. Cao, K. Liang, F. Fang, Z. Fang, T. Li, and K. Gao, Phys. Rev. A **102**, 050802 (2020).
10. M. Takamoto, I. Ushijima, N. Ohmae, T. Yahagi, K. Kokado, H. Shinkai, and H. Katori, Nat. Photonics **14**, 411 (2020).
11. J. Cao, J. Yuan, S. Wang, P. Zhang, Y. Yuan, D. Liu, K. Cui, S. Chao, H. Shu, Y. Lin, S. Cao, Y. Wang, Z. Fang, F. Fang, T. Li, and X. Huang, Appl. Phys. Lett. **120**, 054003 (2022).
12. W. M. Itano, J. C. Bergquist, J. J. Bollinger, J. M. Gilligan, D. J. Heinzen, F. L. Moore, M. G. Raizen, and D. J. Wineland, Phys. Rev. A **47**, 3554 (1993).
13. G. J. Dick, *Proceedings of 19th Annual Precise Time and Time Interval Meeting*, Redondo Beach, California, December 1987, pp. 133–147.
14. D. G. Matei, T. Legero, S. Häfner, C. Grebing, R. Weyrich, W. Zhang, L. Sonderhouse, J. M. Robinson, J. Ye, F. Riehle, and U. Sterr, Phys. Rev. Lett. **118**, 263202 (2017).
15. S. Häfner, S. Herbers, S. Vogt, C. Lisdat, and U. Sterr, Opt. Express **28**, 16407 (2020).
16. S. Häfner, S. Falke, C. Grebing, S. Vogt, T. Legero, M. Merimaa, C. Lisdat, and U. Sterr, Opt. Lett. **40**, 2112 (2015).
17. G. D. Cole, W. Zhang, M. J. Martin, J. Ye, and M. Aspelmeyer, Nat. Photonics **7**, 644 (2013).
18. J. Dickmann and S. Kroker, Phys. Rev. D **98**, 082003 (2018).
19. S. Amairi ep Pyka, "A long optical cavity for sub-Hertz laser spectroscopy," Ph.D. thesis (Gottfried Wilhelm Leibniz Universität Hannover, 2014).
20. D. R. Leibrandt, J. C. Bergquist, and T. Rosenband, Phys. Rev. A **87**, 023829 (2013).
21. S. Webster and P. Gill, Opt. Lett. **36**, 3572 (2011).
22. X. Chen, Y. Jiang, B. Li, H. Yu, H. Jiang, T. Wang, Y. Yao, and L. Ma, Chin. Opt. Lett. **18**, 030201 (2020).
23. B. Argence, E. Prevost, T. Lévèque, R. L. Goff, S. Bize, P. Lemonde, and G. Santarelli, Opt. Express **20**, 25409 (2012).
24. Q.-F. Chen, A. Nevsky, M. Cardace, S. Schiller, T. Legero, S. Häfner, A. Uhde, and U. Sterr, Rev. Sci. Instrum. **85**, 113107 (2014).
25. T. Legero, T. Kessler, and U. Sterr, J. Opt. Soc. Am. B **27**, 914 (2010).
26. G. D. Cole, W. Zhang, B. J. Bjork, D. Follman, P. Heu, C. Deutsch, L. Sonderhouse, J. Robinson, C. Franz, A. Alexandrovski, M. Notcutt, O. H. Heckl, J. Ye, and M. Aspelmeyer, Optica **3**, 647 (2016).
27. S. Herbers, "Transportable ultra-stable laser system with an instability down to $10^{-16}$," Ph.D. thesis, (Gottfried Wilhelm Leibniz Universität Hannover, 2021). Typing error in equation A.28 and A.29: $\exp(-\zeta 2/2)$ instead of $\exp(-\zeta 2/\zeta_{sb,ct})$ and second $\gamma_1$ in equation A.29 must be a $\gamma_2$.
28. J. Ye, J.-L. Peng, R. J. Jones, K. W. Holman, J. L. Hall, D. J. Jones, S. A. Diddams, J. Kitching, S. Bize, J. C. Bergquist, L. W. Hollberg, L. Robertsson, and L.-S. Ma, J. Opt. Soc. Am. B **20**, 1459 (2003).
29. N. C. Wong and J. L. Hall, J. Opt. Soc. Am. B **2**, 1527 (1985).
30. S. Herbers, "Underlying values of Figure 1," figshare (2022), https://doi.org/10.6084/m9.figshare.20337507.
31. A. Farsi, M. Siciliani de Cumis, F. Marino, and F. Marin, J. Appl. Phys. **111**, 043101 (2012).
32. S. Herbers, "Underlying values of Figure 2," figshare (2022), https://doi.org/10.6084/m9.figshare.20337504.
33. T. Chalermsongsak, E. D. Hall, G. D. Cole, D. Follman, F. Seifert, K. Arai, E. K. Gustafson, J. R. Smith, M. Aspelmeyer, and R. X. Adhikari, Metrologia **53**, 860 (2016).
34. U. Sterr, "Frequenzstabilisierungsvorrichtung," German patent DE102 011015489 (16 August 2012).
35. C. Hagemann, "Ultra-stable laser based on a cryogenic single-crystal silicon cavity," Ph.D. thesis (Gottfried Wilhelm Leibniz Universität Hannover, 2013).
36. B. Edlén, Metrologia **2**, 71 (1966).
37. J. Yu, T. Legero, F. Riehle, C. Y. Ma, S. Herbers, D. Nicolodi, D. Kedar, E. Oelker, J. Ye, and U. Sterr, in *2022 Joint Conference of the European Frequency and Time Forum and IEEE International Frequency Control Symposium (EFTF/IFCS)* (2022), pp. 1–3.
38. M. J. Thorpe, D. R. Leibrandt, T. M. Fortier, and T. Rosenband, Opt. Express **18**, 18744 (2010).
39. J. M. Hensley, A. Peters, and S. Chu, Rev. Sci. Instrum. **70**, 2735 (1999).
40. S. Herbers, "Underlying values of Figure S2," figshare (2022), https://doi.org/10.6084/m9.figshare.20337510.
41. S. Herbers, "Underlying values of Figure S3," figshare (2022), https://doi.org/10.6084/m9.figshare.20337513.
42. S. Herbers, "Underlying values of Figure S7," figshare (2022), https://doi.org/10.6084/m9.figshare.21196864.